\documentclass[aps,preprint,showpacs,preprintnumbers,amsmath,amssymb]{revtex4}
\usepackage{amsmath,mathrsfs,amsbsy,color,graphicx,bm,amsthm,amsfonts}
\usepackage{units}
\usepackage{bbm}
\usepackage{times}
\usepackage{dcolumn}
\usepackage{mathrsfs}
\usepackage{amsmath,amssymb,epsfig}
\usepackage{amsmath}
\newcommand{\udots}{\mathinner{\mskip1mu\raise1pt\vbox{\kern7pt\hbox{.}}
\mskip2mu\raise4pt\hbox{.}\mskip2mu\raise7pt\hbox{.}\mskip1mu}}
\begin{document}
\title{Harvesting asymmetric steering via non-identical detectors}
\author{Shu-Min Wu$^1$\footnote{smwu@lnnu.edu.cn}, Rui-Di Wang$^1$,   Xiao-Li Huang$^1$\footnote{ huangxiaoli1982@foxmail.com }, Zejun Wang$^2$ \footnote{zejunwangcz@foxmail.com (corresponding author)}}
\affiliation{$^1$  Department of Physics, Liaoning Normal University, Dalian 116029, China \\
$^2$  Department of Physics, Changzhi University, Changzhi, 046011, China
}


\begin{abstract}
We investigate asymmetric steering harvesting phenomenon  involving two non-identical inertial detectors with different energy gaps, which interact locally with vacuum massless scalar fields. Our study assumes that the energy gap of detector $B$ exceeds that of detector $A$. It is shown that $A\rightarrow B$ steerability  is bigger that $B\rightarrow A$ steerability, implying that  the observer with a small energy gap has more stronger steerability than the other one. We find that the energy gap difference can enlarge the harvesting-achievable range of $A\rightarrow B$ steering, while it can also narrow the harvesting-achievable range of $B\rightarrow A$ steering at the same time. In addition, the maximal steering asymmetry indicates the transformation between two-way steering and one-way steering in some cases, showing that $B\rightarrow A$ steering suffers ``sudden death" at the point of this parameter. These results suggest that asymmetric steering exhibits richer and more interesting properties than quantum entanglement harvested from vacuum quantum field.
\end{abstract}

\vspace*{0.5cm}
 \pacs{04.70.Dy, 03.65.Ud,04.62.+v }
\maketitle
\section{Introduction}
Einstein-Podolsky-Rosen (EPR) steering is a quantum phenomenon that demonstrates nonlocal correlations between entangled systems, commonly referred to as quantum steering. It describes the ability of one observer, known as Alice, to nonlocally influence the state of another observer, known as Bob, through local measurements \cite{L1,L2,L3,L4}. Originally introduced by Schr\"{o}dinger, the concept of quantum steering was further explored by Einstein, Podolsky, and Rosen (EPR) in their seminal 1935 paper, where it became central to the EPR paradox \cite{L5,L6,L7}.
Since the formal redefinition of the steering by Wiseman $et$ $al.$  \cite{L17}, the concept of quantum steering has attracted considerable interest from researchers
\cite{qvom1,qvom2,qvom3,LQE7,L8,L9,L10,L11,L12,L13,L14,L15,RTYR4,L16}. Subsequently, EPR steering is recognized as a form of nonlocality that is stronger than quantum entanglement but weaker than Bell nonlocality \cite{L8,L9}. A key characteristic that distinguishes EPR steering from other types of nonlocality is its intrinsic asymmetry \cite{L10}, meaning that Alice can steer Bob's state, but the reverse is not possible \cite{L11,L12,L13,L14,L15,RTYR4}.  In recent years, there has been a growing focus on both the theoretical and practical implications of quantum steering, particularly in the fields of quantum metrology, quantum computing, quantum secret sharing, and quantum networks \cite{L18,L19,L20,L21,L22,L23,L24,L25,L26,L27,L28}.

In recent years, the study of quantum entanglement and coherence in quantum field theory has made great progress \cite{tjm1,tjm2,tjm3,tjm4,tjm5,tjm6,tjm7,tjm8,tjm9,tjm10,tjm11,tjm12}, such as in the study of black hole entropy and the anti-de Sitter/conformal field theory (AdS/CFT) correspondence \cite{L29,L30,L31}.
It has been shown within the framework of algebraic quantum field theory that the vacuum state of a free quantum field exhibits entanglement, as evidenced by its ability to maximally violate Bell's inequalities \cite{L32,L33}.
It is interesting to note that this vacuum entanglement can be extracted by locally coupling of an initially uncorrelated pair of first-quantized particle detectors to the vacuum field over a finite period of time \cite{L34,L35,L36}. This extraction process has been further operationalized using a protocol based on the Unruh-DeWitt (UDW) detector model, which involves a two-level system \cite{L37,L38,L39,WAX1,L40}. This phenomenon is now commonly referred to as entanglement harvesting. The phenomenon of entanglement harvesting has garnered significant attention and has been investigated under various conditions, including scenarios involving detectors in  the presence of perfectly reflecting boundaries, accelerated motion, and  several aspects of spacetime \cite{L40,L41,L42,L43,L44,L45,L46,L47,L48,L49,QW1,QW2,QW3,QW4,QW5,QW6}.
However, the phenomenon of asymmetric steering harvesting  has not yet received attention. Given that quantum steering represents a crucial quantum resource, it is essential to investigate the process of steering harvested from the vacuum quantum field. This need serves as one of the primary motivations for our research. Unlike quantum entanglement, quantum steering is inherently asymmetric and exhibits a broader range of characteristics, including two-way steering, one-way steering, and no-way steering. Consequently, another key motivation for our study is to explore the asymmetric properties of steering harvesting  for non-identical detectors.

In this paper, based on these two motivations, we investigate  the phenomenon of steering harvesting for two non-identical detectors $A$ and $B$.  For the sake of clarity in our discussion, we assume that the energy gap of detector $B$ is larger than that of detector $A$.
We will mainly consider how the presence of energy gap differences affect the steerabilities  $A\rightarrow B$ and $B\rightarrow A$ harvested, the range of harvesting-achievable separation, and the steering asymmetry.  Specifically, we address three questions: (i) whether non-identical detectors could harvest more steerabilities  $A\rightarrow B$ and $B\rightarrow A$ than identical detectors; (ii) how the energy gap difference between the two detectors affects the harvesting-achievable range of steering in the $A\rightarrow B$ and $B\rightarrow A$ directions; (iii) how the energy gap difference influences the steering asymmetry and the transition between  two-way steering and one-way steering. By exploring these questions, we aim to uncover the novel  properties associated with harvesting asymmetric steering from the vacuum quantum field.

The paper is organized as follows. In Sec. II, we briefly introduce the quantification of quantum steering for  the X-state. In Sec. III, we discuss the Unruh-DeWitt model for non-identical detectors.  In Sec. IV, we study the properties of asymmetric steering harvesting for two inertial non-identical detectors with different energy gaps.  Finally, we end with a brief conclusion in the last section.

\section{Quantification of quantum steering for X-state}
Quantum steering is one of the representative forms of quantum correlation. Initially, we briefly introduce the definition and measurement of quantum steering. By considering a bipartite quantum system involving two particles, Alice and Bob, existing in a joint quantum state, we consider the density matrix of the X-state $\rho_{AB}$ as
\begin{eqnarray}\label{S1}
\rho_{AB}=\left(\!\!\begin{array}{cccccccc}
\rho_{11} & 0 & 0 & \rho_{14}\\
0 & \rho_{22} & \rho_{23} & 0\\
0 & \rho_{32} & \rho_{33} & 0\\
\rho_{41} & 0 & 0 & \rho_{44}\\
\end{array}\!\!\right),
\end{eqnarray}
where $\rho_{ij}$ is an element of the matrix and satisfies $\rho_{ij}^*=\rho_{ji}$. It is noteworthy that concurrence serves as a robust metric for discerning quantum entanglement in bipartite states, denoted as \cite{L52}
\begin{eqnarray}\label{S2}
C(\rho_{AB})=2\max\{0, |\rho_{14}|-\sqrt{\rho_{22}\rho_{33}}, |\rho_{23}|-\sqrt{\rho_{11}\rho_{44}}\}.
\end{eqnarray}
Then, for the X-state $\rho_{AB}$ shared by Alice and Bob, the presence of steering from Bob to Alice can be observed through the entanglement of the density matrix $\tau_{AB}$, defined as  \cite{L53,L54}
\begin{eqnarray}\label{S3}
\tau_{AB}=\frac{\rho_{AB}}{\sqrt{3}}+\frac{3-\sqrt{3}}{3}(\rho_{A}\otimes\frac{I}{2}),
\end{eqnarray}
where $\rho_{A}$ is the reduced density matrix with respect to Alice, given by $\rho_{A}=Tr_{B}(\rho_{AB})$, and $I$ denotes the two-dimensional identity matrix. Likewise, the steering from Alice to Bob can be witnessed if the state $\tau_{BA}$
defined as
\begin{eqnarray}\label{S4}
\tau_{BA}=\frac{\rho_{AB}}{\sqrt{3}}+\frac{3-\sqrt{3}}{3}(\frac{I}{2}\otimes\rho_{B}),
\end{eqnarray}
is entangled, where $\rho_{B}=Tr_{A}(\rho_{AB})$ is reduced density matrix of Bob.

Using Eqs.(\ref{S1}) and (\ref{S3}), the matrix $\tau_{AB}$ of Eq.(\ref{S1}) can be specifically written as
\begin{eqnarray}
\tau_{AB}=\left(\!\!\begin{array}{cccccccc}
\frac{\sqrt{3}}{3}\rho_{11}+f & 0 & 0 & \frac{\sqrt{3}}{3}\rho_{14}\\
0 & \frac{\sqrt{3}}{3}\rho_{22}+f & \frac{\sqrt{3}}{3}\rho_{23} & 0\\
0 & \frac{\sqrt{3}}{3}\rho_{32} & \frac{\sqrt{3}}{3}\rho_{33}+h & 0\\
\frac{\sqrt{3}}{3}\rho_{41} & 0 & 0 & \frac{\sqrt{3}}{3}\rho_{44}+h\\
\end{array}\!\!\right),
\end{eqnarray}
with $f=\frac{(3-\sqrt{3})}{6}(\rho_{11}+\rho_{22})$ and $h=\frac{(3-\sqrt{3})}{6}(\rho_{33}+\rho_{44})$. By employing Eq.(\ref{S2}), the state $\tau_{AB}$ is entangled  if either of the following inequalities holds
\begin{eqnarray}
|\rho_{14}|>\sqrt{J_{a}-J_{b}},
\end{eqnarray}
\begin{eqnarray}
|\rho_{23}|>\sqrt{J_{c}-J_{b}},
\end{eqnarray}
 where
\begin{gather}
J_{a}=\frac{2-\sqrt{3}}{2}\rho_{11}\rho_{44}+\frac{2+\sqrt{3}}{2}\rho_{22}\rho_{33}+\frac{1}{4}(\rho_{11}+\rho_{44})(\rho_{22}+\rho_{33}),\nonumber\\
J_{b}=\frac{1}{4}(\rho_{11}-\rho_{44})(\rho_{22}-\rho_{33}),\nonumber\\
J_{c}=\frac{2+\sqrt{3}}{2}\rho_{11}\rho_{44}+\frac{2-\sqrt{3}}{2}\rho_{22}\rho_{33}+\frac{1}{4}(\rho_{11}+\rho_{44})(\rho_{22}+\rho_{33}).
\end{gather}
The steering from Alice to Bob can be observed analogously to the steering from Bob to Alice, verified by satisfying one of the inequalities
\begin{eqnarray}
|\rho_{14}|>\sqrt{J_{a}+J_{b}},
\end{eqnarray}
or
\begin{eqnarray}
|\rho_{23}|>\sqrt{J_{c}+J_{b}}.
\end{eqnarray}
Based on these criteria, the steerability from Bob to Alice $S^{B\rightarrow A}$ and from Alice to Bob $S^{A\rightarrow B}$ can be quantified as
\begin{eqnarray}\label{TBA}
S^{B\rightarrow A}=\max\{0, |\rho_{14}|-\sqrt{J_{a}-J_{b}}, |\rho_{23}|-\sqrt{J_{c}-J_{b}}\},
\end{eqnarray}
\begin{eqnarray}\label{TAB}
S^{A\rightarrow B}=\max\{0, |\rho_{14}|-\sqrt{J_{a}+J_{b}}, |\rho_{23}|-\sqrt{J_{c}+J_{b}}\}.
\end{eqnarray}

\section{Unruh-DeWitt model  for non-identical detectors}
We consider a pair of two-level detectors labeled by $A$ and $B$, characterized by the ground state $|0\rangle_{D}$ and the excited state $|1\rangle_{D}$. These detectors interact locally with a massless scalar field $\phi[\mathsf{x}_D(\tau)]$ ($D\in\{A, B\}$) in vacuum. Then, the Hamiltonian of the interaction between the detectors and the field  can be expressed as
\begin{eqnarray}\label{S11}
H_D(t) &= \lambda \chi\! \left(t \right)\Big(e^{ i\Omega_{D} \tau} \sigma^+  +  e^{- i\Omega_{D} \tau}\sigma^- \Big) \otimes  \phi\left[\mathsf{x}_D(\tau)\right],
\end{eqnarray}
where the operators $\sigma^+ =|{1_D}\rangle\langle{0_D}|$ and $\sigma^-=|{0_D}\rangle\langle{1_D}|$ are the SU(2) ladder operators acting on the detector Hilbert space, $\chi(\tau) = \exp[-\tau^2/2\sigma^2]$ is a Gaussian switching function of the parameter $\sigma$ that controls the duration of the interaction, $\mathsf{x}_D(\tau)$ is the spacetime trajectory of the detector, parameterized by its proper time $\tau$, $\lambda$ is the coupling strength, and $\Omega_D$ is the energy gap of the detector.

Initially, we consider two UDW detectors prepared in their respective ground states, denoted as $|{0} \rangle_A|{0} \rangle_B$, and the field state is in the vacuum state $|0\rangle_M$. Thus, the initial state of the composite system is  $|{\Psi_i}\rangle = |{0} \rangle_A|{0} \rangle_B|{0}\rangle_M$.
The time evolution of the quantum system can be obtained by using the Hamiltonian of Eq.(\ref{S11}),
\begin{eqnarray}\label{S12}
|{\Psi_f}\rangle = \mathcal{T} \exp\left[  -i \int_{\mathbb{R}}  dt\, \left(\frac{d \tau_A}{dt}H_A(\tau_A) + \frac{d \tau_B}{dt} H_B(\tau_B) \right) \right]|{\Psi_i}\rangle,
\end{eqnarray}
where $\mathcal{T}$ is the time ordering operator, $H_A( t)$ and $H_B(t)$ are specified in Eq.(\ref{S11}), and $t$ is the coordinate time with respect to which the vacuum state of the field is defined. The final reduced state of the detectors can be derived in the basis $\{ |{0_A 0_B}\rangle, |{0_A 1_B}\rangle, |{1_A 0_B}\rangle, |{1_A 1_B}\rangle \}$ by tracing out the field degrees of freedom in Eq.(\ref{S12}) \cite{L42,L55,L56}
\begin{eqnarray}\label{AB}
\rho_{AB} :=\rm{Tr}_{\phi}(\rm U|\Psi_{f}\rangle\langle\Psi_{f}|\rm U^{\dag} )
= \begin{pmatrix}
1 - P_A-P_B  & 0 & 0 & X \\
0 & P_B  & C & 0 \\
0 & C^* & P_A & 0 \\
X^* & 0 & 0 & 0
\end{pmatrix} + \mathcal{O}\!\left(\lambda^4\right),
\end{eqnarray}
where
\begin{eqnarray}\label{P1}
P_{D}:=\lambda^{2}\int\int{d\tau d\tau'\chi(\tau)\chi(\tau')e^{-i\Omega_D(\tau-\tau')}W(\mathsf{x}_{D}(\tau),\mathsf{x}_{D}(\tau'))}\ \quad   D\in\{A,B\},
\end{eqnarray}
\begin{eqnarray}
C:=\lambda^{2}\int\int{d\tau d\tau'\chi(\tau)\chi(\tau')e^{-i(\Omega_A\tau-\Omega_B\tau')}W(\mathsf{x}_{A}(\tau),\mathsf{x}_{B}(\tau'))},
\end{eqnarray}
\begin{eqnarray}
X:&=&-\lambda^{2}\int\int{d\tau d\tau'\chi(\tau)\chi(\tau')e^{-i(\Omega_A\tau+\Omega_B\tau')}}\nonumber\\
&&[\theta(\tau'-\tau)W(\mathsf{x}_{A}(\tau),\mathsf{x}_{B}(\tau'))+\theta(\tau-\tau')W(\mathsf{x}_{B}(\tau'),\mathsf{x}_{A}(\tau))]
\end{eqnarray}
with the vacuum Wightman function $W(\mathsf{x},\mathsf{x}')= \langle 0|_M\phi(\mathsf{x}) \phi(\mathsf{x}') | 0\rangle_M$  obtained by detailed calculations and the Heaviside's step function $\theta(\tau)$. Note that $P_D$ is called the transition probability of detector $D$, and the quantities $C$ and $X$ characterize the correlation terms \cite{L56}.

It is noteworthy that prior investigations have predominantly focused on identical detectors, indicating that the relationship between the two detectors is symmetrical.
However, our model considers non-identical detectors with different energy gaps, meaning that the relationship between the two detectors is asymmetric. Based on this model, harvesting  steering  from vacuum quantum fields may be  asymmetric.
For the sake of clarity, we assume without loss of generality that detector $A$ possesses a relatively smaller energy gap compared to detector $B$, i.e., $\Delta\Omega=\Omega_B-\Omega_A\geq0$ throughout the paper. The Wightman function of a massless scalar field in the four-dimensional Minkowski spacetime is represented by \cite{L56-1,L57,L58}
\begin{equation}\label{WXX}
W(\mathsf{x},\mathsf{x'})=-\frac{1}{4 \pi^2}\frac{1}{(\tau-\tau'-i \epsilon)^2-|\mathbf{x}-\mathbf{x'}|^2}.
\end{equation}
Combining Eqs.(\ref{P1}) and (\ref{WXX}), one can obtain the transition probability as \cite{L39}
\begin{equation}
P_D = \frac{\lambda^2}{4 \pi} \left[ e^{-\sigma^2 \Omega_D^2} - \sqrt{\pi} \Omega_D\sigma \rm{Erfc}(\sigma\Omega_D)\right],\quad D\in\{A,B\}.
\end{equation}
Similarly, the correlation terms $C$ and $X$ in this case can also be computed as
\begin{eqnarray}
X=i\frac{\lambda^2\sigma}{8\sqrt{\pi}L}e^{-\frac{\sigma^4(2\Omega_A+\Delta\Omega)^2+L^2}{4\sigma^2}}&&\bigg[e^{\frac{i\Delta\Omega L}{2}}\rm{Erfi}(\frac{L-i\sigma^2\Delta\Omega}{2\sigma})\notag\\
&&+e^{-\frac{i\Delta\Omega L}{2}}\rm{Erfi}(\frac{L+i\sigma^2\Delta\Omega}{2\sigma})+2i\cos(\frac{\Delta\Omega L}{2})\bigg],\
\end{eqnarray}
\begin{eqnarray}
C= \frac{\lambda^2\sigma}{4\sqrt{\pi}L}e^{-\frac{L^2+\Delta\Omega^2\sigma^4}{4\sigma^2}}&&\bigg(\rm{Im}\big[e^{\frac{i(2\Omega_A+\Delta\Omega)L}{2}}\rm{Erf}(\frac{iL+2\Omega_A\sigma^2+\Delta\Omega\sigma^2}{2\sigma})\big]\notag\\
&&-\sin{\frac{(2\Omega_A+\Delta\Omega)L}{2}}\bigg).
\end{eqnarray}

\section{Quantum steering  harvesting for non-identical detectors}
In this section, we will consider how the presence of energy gap differences affect the steering asymmetry,  the range of harvesting-achievable separation, and the amount of  steering harvested.  Employing Eqs.(\ref{TBA}), (\ref{TAB}), and (\ref{AB}), we can obtain the analytic expressions of the steering $S^{B\rightarrow A}$ and $S^{A\rightarrow B}$ as
\begin{equation}
\begin{aligned}\label{lkj1}
S^{B\rightarrow A}=\max \bigg\{0, &|X|-\sqrt{\frac{1+\sqrt{3}}{2}P_A P_B+\frac{1}{2}P_A-\frac{1}{2}{P_A}^2},\\
&|C|-\sqrt{\frac{1-\sqrt{3}}{2}P_A P_B+\frac{1}{2}P_A-\frac{1}{2}{P_A}^2}\bigg\},
\end{aligned}
\end{equation}
and
\begin{equation}
\begin{aligned}\label{lkj2}
S^{A\rightarrow B}=\max \bigg\{0, &|X|-\sqrt{\frac{1+\sqrt{3}}{2}P_A P_B+\frac{1}{2}P_B-\frac{1}{2}{P_B}^2},\\
&|C|-\sqrt{\frac{1-\sqrt{3}}{2}P_A P_B+\frac{1}{2}P_B-\frac{1}{2}{P_B}^2}\bigg\}.\
\end{aligned}
\end{equation}
To further investigate whether the steering $S^{B\rightarrow A}$ and $S^{A\rightarrow B}$ is symmetric, we also calculate the difference between $S^{B\rightarrow A}$ and $S^{A\rightarrow B}$, referred to as the steering asymmetry
\begin{equation}
S^\Delta_{AB}=|S^{B\rightarrow A}-S^{A\rightarrow B}|.
\end{equation}

\begin{figure}
\begin{minipage}[t]{0.5\linewidth}
\centering
\includegraphics[width=3.0in,height=5.2cm]{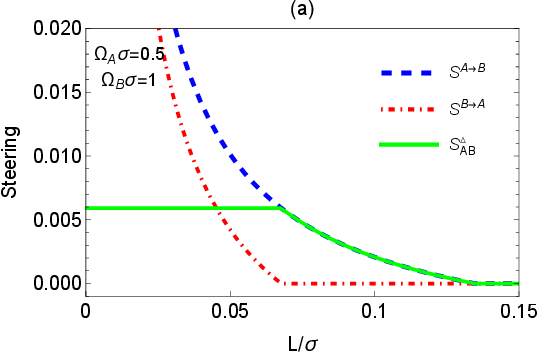}
\label{fig1a}
\end{minipage}%
\begin{minipage}[t]{0.5\linewidth}
\centering
\includegraphics[width=3.0in,height=5.2cm]{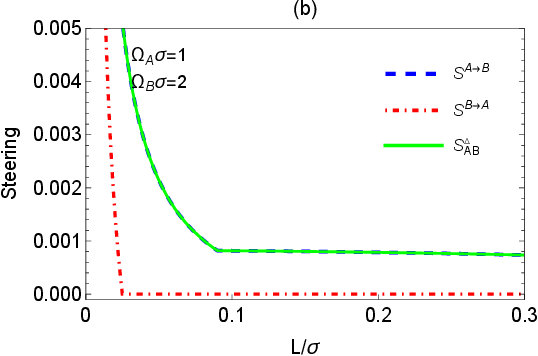}
\label{fig1b}
\end{minipage}%

\caption{Quantum steering $S^{B\rightarrow A}$, $S^{A\rightarrow B}$, and the steering asymmetry $S^\Delta_{AB}$ of two non-identical detectors as a function of the detector separation $L/\sigma$ for the fixed energy gap difference $\Delta\Omega/\Omega_A=1$. The coupling constant is set to $\lambda=0.1$.}
\label{Fig.1}
\end{figure}

In Fig.\ref{Fig.1}, we plot quantum steering $S^{A\rightarrow B}$,  $S^{B\rightarrow A}$,  as well
as the steering asymmetry $S^\Delta_{AB}$ as a function of the detector separation $L/\sigma$ when the energy gap difference $\Delta\Omega/\Omega_A$ is fixed to 1 with different values of $\Omega_A\sigma$.
We find that the steering  first monotonically decreases and then suffers ``sudden death" with increasing detector separation $L/\sigma$.
We also find that  the steerability $S^{A\rightarrow B}$  is always bigger than the steerability $S^{B\rightarrow A}$   for non-identical detectors, meaning that the observer who has a small energy gap  has more stronger steerability than the other one. This represents that harvesting steering is asymmetrical because of the energy gap difference. Naturally, quantum steering $S^{B\rightarrow A}$ suffers ``sudden death" earlier than quantum steering $S^{A\rightarrow B}$.
In other words, the range of harvesting-achievable separation of the steering $S^{A\rightarrow B}$  is wider than that of the steering $S^{B\rightarrow A}$.
The ``sudden death" of quantum steering $S^{B\rightarrow A}$  indicates the system $\rho_{AB}$ is currently experiencing a transformation from two-way steering to one-way steering. In addition,  the ``sudden death" of quantum steering $S^{A\rightarrow B}$ implies that  the system $\rho_{AB}$ are transitioning from one-way steering to no-way steering.

\begin{figure}[htbp]
\begin{minipage}[t]{0.5\linewidth}
\centering
\includegraphics[width=3.0in,height=5.2cm]{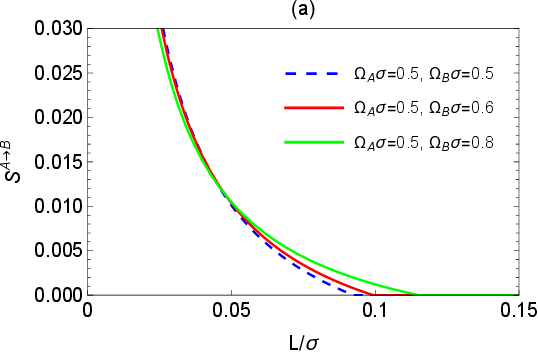}
\end{minipage}%
\begin{minipage}[t]{0.5\linewidth}
\centering
\includegraphics[width=3.0in,height=5.2cm]{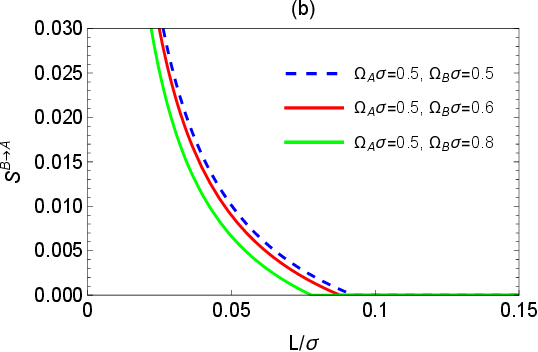}
\end{minipage}%

\begin{minipage}[t]{0.5\linewidth}
\centering
\includegraphics[width=3.0in,height=5.2cm]{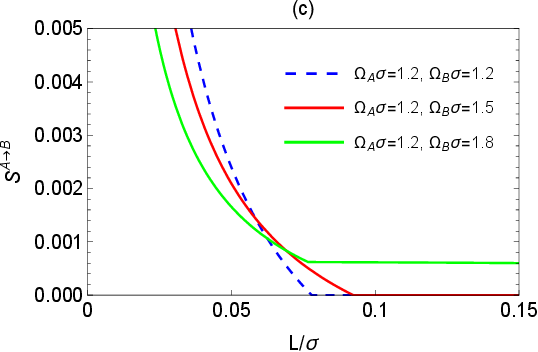}
\end{minipage}%
\begin{minipage}[t]{0.5\linewidth}
\centering
\includegraphics[width=3.0in,height=5.2cm]{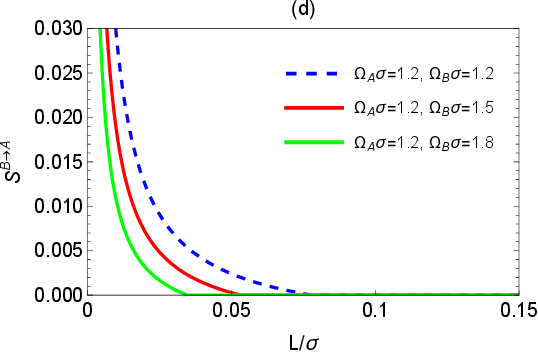}
\end{minipage}%
\caption{Quantum steering $S^{A\rightarrow B}$ and $S^{B\rightarrow A}$ between two non-identical detectors as a function of the detector separation $L/\sigma$ for $\Delta\Omega/\Omega_A=\{0,0.2,0.6\}$ with $\Omega_A\sigma=0.5$ in (a) and (b), $\Delta\Omega/\Omega_A=\{0,0.25,0.5\}$ with $\Omega_A\sigma=1.2$ in (c) and (d). The coupling constant is set to $\lambda=0.1$.}\label{Fig.2}
\end{figure}

In Fig.\ref{Fig.2}, we describe how quantum steering $S^{A\rightarrow B}$ and $S^{B\rightarrow A}$ vary as a function of the detector separation $L/\sigma$ with different values of $\Omega_A\sigma$ for unfixed $\Delta\Omega/\Omega_A$. It is noteworthy that, regardless of the value of $\Delta\Omega/\Omega_A$, the harvested steerability decreases with increasing detector separation $L/\sigma$, which is consistent with existing results  on harvested steerability for the  identical detectors $\Delta\Omega=0$. However, it is important to emphasize that, the energy
gap difference does impact the amount of quantum steering harvested and the range of harvesting-achievable separation. From Fig.\ref{Fig.1} (a) and (c), we find that, for the smaller detector separation $L/\sigma$, increasing energy gap difference $\Delta\Omega/\Omega_A$ may be detrimental to the steerability $S^{A\rightarrow B}$, implying that the energy gap difference $\Delta\Omega/\Omega_A$ has a negative effect on quantum steering $S^{A\rightarrow B}$. For the bigger detector separation $L/\sigma$, increasing energy gap difference $\Delta\Omega/\Omega_A$ is beneficial to improving the steerability $S^{A\rightarrow B}$, which means that  non-identical detectors with an energy gap difference inevitably harvest more steerability $S^{A\rightarrow B}$ via locally interacting with vacuum fields than
identical detectors.  In this sense, the presence of the energy gap difference $\Delta\Omega/\Omega_A$ can have a positive effect on quantum steering $S^{A\rightarrow B}$. However, increasing energy gap difference $\Delta\Omega/\Omega_A$ consistently reduces the steerability $S^{B\rightarrow A}$,
indicating that non-identical detectors  are bound to harvest less steerability $S^{B\rightarrow A}$ from vacuum fields than identical detectors. It is interesting to discover that  the energy gap difference $\Delta\Omega/\Omega_A$ can enlarge the harvesting-achievable range of quantum steering $S^{A\rightarrow B}$, while  the energy gap difference $\Delta\Omega/\Omega_A$ can narrow  the harvesting-achievable range of quantum steering $S^{B\rightarrow A}$. Different from quantum entanglement, asymmetric steering exhibits richer and more interesting properties for non-identical detectors. It is worth noting that  $S^{A\rightarrow B}$ also suffers from sudden death with the growth of the $L/\sigma$, both for the case of  $\Omega_A\sigma=1$ and $\Omega_B\sigma=2$ in Fig.\ref{Fig.1}(b) and $\Omega_A\sigma=1.2$ and $\Omega_B\sigma=1.8$ in Fig.\ref{Fig.2}(c). To make the figures clearer, we restrict the plots to a smaller range of $L/\sigma$.

\begin{figure}
\begin{minipage}[t]{0.5\linewidth}
\centering
\includegraphics[width=3.0in,height=5.2cm]{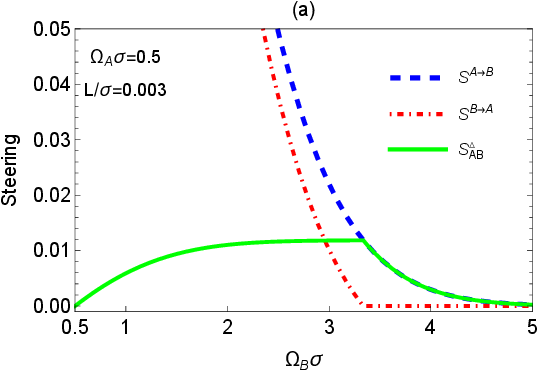}
\label{fig3a}
\end{minipage}%
\begin{minipage}[t]{0.5\linewidth}
\centering
\includegraphics[width=3.0in,height=5.2cm]{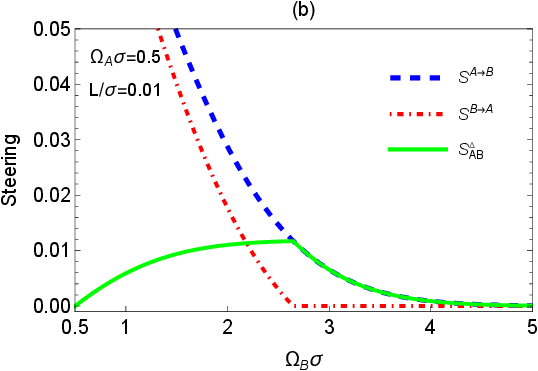}
\label{fig3b}
\end{minipage}%

\begin{minipage}[t]{0.5\linewidth}
\centering
\includegraphics[width=3.0in,height=5.2cm]{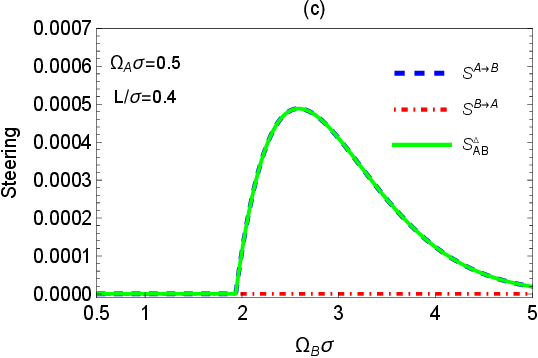}
\label{fig3c}
\end{minipage}%
\begin{minipage}[t]{0.5\linewidth}
\centering
\includegraphics[width=3.0in,height=5.2cm]{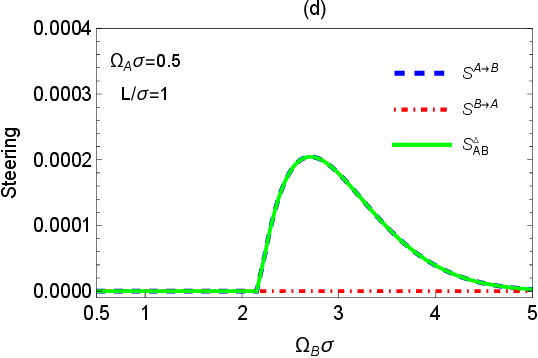}
\label{fig3d}
\end{minipage}%

\caption{Quantum steering $S^{A\rightarrow B}$, $S^{B\rightarrow A}$, and the steering asymmetry $S^\Delta_{AB}$ of two non-identical detectors as a function of the energy gap $\Omega_B\sigma$ with $\Omega_A\sigma=0.5$ for the smaller $L/\sigma=\{0.003, 0.01\}$  in (a) and (b), the larger $L/\sigma=\{0.4, 1\}$ in (c) and (d), respectively.  The coupling constant is set to $\lambda=0.1$.}
\label{Fig.3}
\end{figure}

In Fig.\ref{Fig.3}, we illustrate the variation of the steering $S^{A\rightarrow B}$, $S^{B\rightarrow A}$,  and the steering asymmetry $S^\Delta_{AB}$ with the energy gap $\Omega_B\sigma$ for different values of $L/\sigma$.
It should be stated that an increase in variable $\Omega_B$ is also theoretically equivalent to an increase in $\Delta\Omega/\Omega_A$ ($\Delta\Omega/\Omega_A \propto \Omega_B$). From Fig.\ref{Fig.3}, we can see that  the steering asymmetry for the  identical detectors ($\Delta\Omega=0$) is always equal to zero, while  the steering asymmetry for non-identical detectors ($\Delta\Omega\neq 0$) may be harvested from the vacuum field. This result indicates that non-identical detectors cause the steering asymmetry  harvested from the vacuum field. From Fig.\ref{Fig.3}(a) and (b), we find that the maximal steering asymmetry $S^\Delta_{AB}$  indicates a transition between  two-way steerability to one-way steerability, meaning that the parameters attaining the peaks of steering asymmetry $S^\Delta_{AB}$  are obtained when the steering $S^{B\rightarrow A}$ experiences ``sudden death".
From Fig.\ref{Fig.3}(c) and (d), we find that non-identical detectors can create quantum steering $S^{A\rightarrow B}$ and cannot generate quantum steering $S^{B\rightarrow A}$ harvested from the vacuum field, reflecting that the energy gap difference $\Delta\Omega/\Omega_A$ can broaden the harvesting-achievable range for harvested one-way steering.  As shown in Fig.\ref{Fig.2} and Fig.\ref{Fig.3}, quantum steering harvested by time-like separated detectors decreases significantly with increasing energy gap difference when inter-detector separation is very small ($L/\sigma\ll1$).
For very small separation $L/\sigma\ll1$,   Eqs.(\ref{lkj1}) and (\ref{lkj2}) show that the correlation terms dominate quantum steering, while the transition probability term becomes negligible \cite{L39,L57,L58}. As a result, quantum steering decreases with increasing energy gap difference. However, for large separation ($L/\sigma \gg 1$), the correlation terms are also suppressed, and the transition probability can no longer be ignored. Numerical results show that, in this case, quantum steering can hardly be extracted effectively from the vacuum.

\begin{figure}
\begin{minipage}[t]{0.5\linewidth}
\centering
\includegraphics[width=3.0in,height=5.2cm]{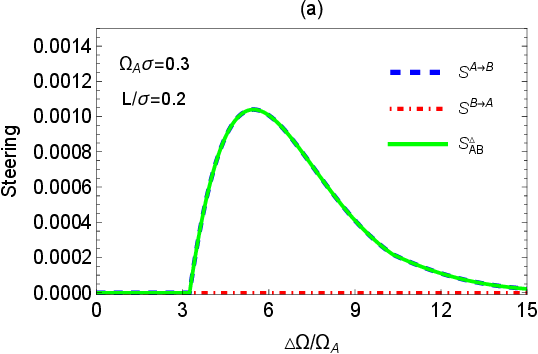}
\label{fig4a}
\end{minipage}%
\begin{minipage}[t]{0.5\linewidth}
\centering
\includegraphics[width=3.0in,height=5.2cm]{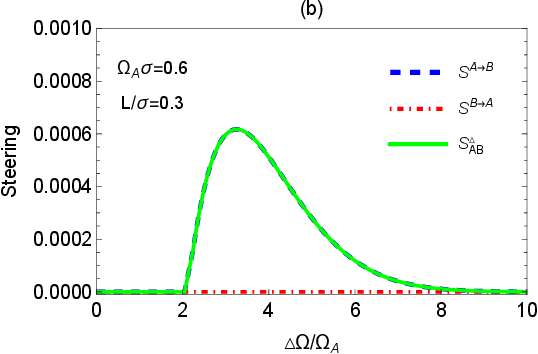}
\label{fig4b}
\end{minipage}%

\begin{minipage}[t]{0.5\linewidth}
\centering
\includegraphics[width=3.0in,height=5.2cm]{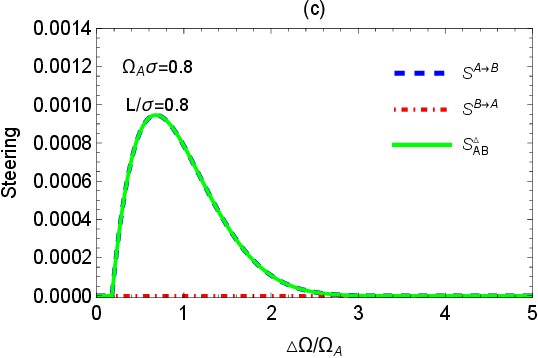}
\label{fig4c}
\end{minipage}%
\begin{minipage}[t]{0.5\linewidth}
\centering
\includegraphics[width=3.0in,height=5.2cm]{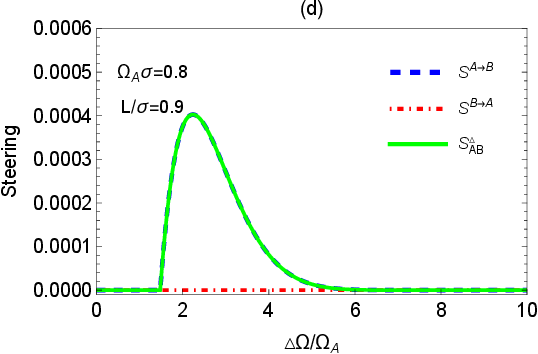}
\label{fig4d}
\end{minipage}%

\caption{Quantum steering $S^{A\rightarrow B}$, $S^{B\rightarrow A}$, and the steering asymmetry $S^\Delta_{AB}$ of two non-identical detectors as a function of the energy gap difference $\Delta\Omega/\Omega_A$. The coupling constant is set to $\lambda=0.1$.}
\label{Fig.4}
\end{figure}

In Fig.\ref{Fig.4}, we describe how steering $S^{A\rightarrow B}$, $S^{B\rightarrow A}$, and the steering asymmetry $S^\Delta_{AB}$  vary with the energy gap differences $\Delta\Omega/\Omega_A$ for different values of $L/\sigma$ and $\Omega_A\sigma$. We can clearly find that $S^{A\rightarrow B}$ can be harvested through non-identical detectors, while the reverse steering $S^{B\rightarrow A}$ is strongly suppressed and often vanishes. This obviously indicates that the existence of the energy gap difference $\Delta\Omega/\Omega_A$ can expand the harvesting-achievable range of one-way steering.

\section{Conclusion}
The phenomenon of asymmetric steering harvesting for two non-identical detectors with different energy gaps locally interacting with massless scalar fields  is investigated in fat spacetime. Our analysis assumes that the energy gap of detector $B$ is bigger than that of detector $A$.  We analyze the influence of the energy gap difference between the two detectors on steering harvesting from three perspectives: (i) $A\rightarrow B$ and $B\rightarrow A$ steerabilities; (ii) the range of harvesting-achievable separation; (iii) the amount of steering asymmetry.  It is shown that the steering is always asymmetric and $A\rightarrow B$  steerability is bigger than $B\rightarrow A$ steerability harvested from vacuum quantum field, indicating that  the observer who has a smaller  energy gap has stronger steerability than the other one. An interesting discovery is that  the energy gap difference can narrow the range of harvesting-achievable separation for  $B\rightarrow A$ steering, while it can also broaden the range of harvesting-achievable separation for $A\rightarrow B$ steering, which means that the energy gap difference  can create $A\rightarrow B$ steering harvested from the vacuum field. For some circumstances, the maximal steering asymmetry indicates that the system is experiencing a transformation from two-way steering to one-way steering under the influence of the energy gap difference, showing that $B\rightarrow A$ steering suffers ``sudden death" at the point of this parameter. Our results suggest that the energy gap difference  can manipulate one-way steering and two-way steering harvesting, which can be used to process quantum information tasks. Unlike quantum entanglement, quantum steering possesses intrinsic directionality, giving rise to richer structures such as one-way and two-way steering. It is possible to extract asymmetric quantum steering from the vacuum through  non-identical detectors.
By explicitly comparing the analytical expressions of concurrence (used to quantify entanglement) with those of quantum steering in both directions ($A \rightarrow B$ and $B \rightarrow A$), it is observed that the spatial separation required for harvesting quantum steering is smaller than that for harvesting entanglement. This implies that entanglement  can persist and be extracted under conditions where steering may no longer be accessible.
These results suggest that quantum steering, particularly in asymmetric settings, provides a more unique resource for quantum communication and control in relativistic scenarios.

\begin{acknowledgments}
This work is supported by the National Natural
Science Foundation of China (Grant Nos. 12205133), LJKQZ20222315, JYTMS20231051,  the Special Fund for Basic Scientific Research of Provincial Universities in Liaoning under grant NO. LS2024Q002,  and Scientific and Technological Innovation Programs of Higher Education
Institutions of Shanxi Province, China (2023L327).
\end{acknowledgments}


\end{document}